\pdfoutput=1
\documentclass[aps,prd,amsmath,floats,floatfix,twocolumn,
superscriptaddress,nofootinbib,showpacs]{revtex4-1}

\usepackage[T1]{fontenc}
\usepackage[utf8]{inputenc}
\usepackage{lmodern}

\usepackage{verbatim}

\usepackage[dvipsnames, usenames]{xcolor}
\definecolor{linkcolor}{rgb}{0.0,0.3,0.5}
\usepackage[hypertexnames=false, unicode, colorlinks=true, linkcolor=linkcolor,
citecolor=linkcolor, filecolor=linkcolor,urlcolor=linkcolor,
pdfusetitle]{hyperref}

\usepackage[all]{hypcap}
\usepackage{graphicx}
\usepackage{xspace}
\usepackage{amssymb}
\usepackage{microtype}

\usepackage[english]{babel}
\usepackage{blindtext}

\usepackage[normalem]{ulem} %for \sout
\usepackage{bm} % boldmath
\usepackage{mathrsfs}

\newcommand\eea{\end{eqnarray}}
\newcommand\bea{\begin{eqnarray}}
\newcommand\ea{\end{align}}
\newcommand\ba{\begin{align}}

\newcommand\nn{\nonumber}
\newcommand\ml{\mathscr}

\usepackage{natbib}
\usepackage{graphicx}
\usepackage[dvipsnames]{xcolor}
\usepackage{amsmath}
\begin{document}

\title{Scalar and Gravitational Transient ``Hair'' for Near-Extremal Black Holes}
\newcommand{\UMassDPhy}{\affiliation{Department of Physics,
		University of Massachusetts, Dartmouth, MA 02747, USA}}
\newcommand{\CSCVRUMass}{\affiliation{Center for Scientific Computing and Data Research, University of Massachusetts, Dartmouth, MA 02747, USA}}
\newcommand{\URI}{\affiliation{Department of Physics, 
    University of Rhode Island, Kingston, RI 02881, USA}}    

\author{Kevin Gonz\'alez-Quesada}
\email{kgonzalezquesada@umassd.edu}
\UMassDPhy
\CSCVRUMass

\author{Subir Sabharwal}
\email{subir.sabharwal@gmail.com}
\UMassDPhy
\CSCVRUMass

\author{Gaurav Khanna}
\email{gkhanna@uri.edu}
\URI
\UMassDPhy
\CSCVRUMass

\date{\today}

%==========================================================================
\begin{abstract}
We study the existence and nature of Aretakis ``hair'' and its potentially observable imprint at a finite distance from the horizon (Ori-coefficient) in near-extremal black hole backgrounds. Specifically, we consider the time evolution of horizon penetrating scalar and gravitational perturbations with compact support on near-extremal Reissner-Nordstr{\"o}m (NERN) and Kerr (NEK). We do this by numerically solving the Teukolsky equation and determining the Aretakis charge values on the horizon and at a finite distance from the black hole. We demonstrate that these values are no longer strictly conserved in the non-extremal case; however, their decay rate can be arbitrarily slow as the black hole approaches extremality allowing for the possibility of their observation as a {\em transient} hair. 
\end{abstract}

\maketitle

\section{Introduction}
Black holes are perhaps the most interesting astrophysical objects predicted by general relativity (GR). In contrast to other astrophysical objects like stars, white dwarfs or pulsars, black holes (BH) are an exclusive prediction of GR, in that they do not exist in a Newtonian description of gravitation. They arise as simple static or stationary metric solutions to the Einstein field equations of the geometry outside a massive object. 
Their simplicity derives, in part, from the fact that any BH can be described solely by three parameters -- mass, global charge and spin, irrespective of the details of their formation. This statement is somewhat formally encoded in Bekenstein's so-called ``no-hair'' theorem~\cite{bek72a,bek72b,bek72c}. However,  certain caveats to this theorem have been recently exploited in the context of an extremal BH (EBH) in work completed by Aretakis and others~\cite{are12,aag18,lmrt13,zimm17,lr12,gz18,bk18,bks21}. These papers consider either extremal Reissner-Nordstr{\"o}m (ERN) or extremal Kerr (EK) black holes, and by solving for evolution of a scalar or gravitational field satisfying certain initial conditions on these EBH background, it can be shown that there is a conserved charge on the horizon, $\ml{H}^+$ which is evaluated solely based on the radial derivative of the field value on the horizon at late times. Moreover, this charge is a ``hair'' in the sense that it can also be observed from outside the BH -- this was shown analytically for ERN BHs in Ref.~\cite{aag18} with an exact expression for the horizon hair at null infinity, $\ml{I}^+$ for the spherically symmetric scalar field case. Specifically, it was shown that the horizon charge could be calculated solely by observing the field values at $\ml{I}^+$. In addition, Ori supplemented this work with an expression for the Aretakis charge at finite radial distance from an ERN BH~\cite{ori}. This was generalized via numerical simulations to the Kerr and the gravitational field case in Ref.~\cite{bks21}. The existence of the horizon hair of Aretakis has been verified numerically for ERN and EK BHs at $\ml{H}^+$ as well as at $\ml{I}^+$~\cite{bk18}. Numerical work has also been extended to the Ori results as well as been generalized to the case of extremal Kerr black holes for the scalar and gravitational field cases~\cite{bks21}.\\ %{\bf (Maybe move this to last section on future work?):} In upcoming work, this is generalized to the case of non-spherically symmetric perturbations on EBH backgrounds. 

Even though extremal BHs are an important theoretical and mathematical curiosity, we know from the arguments of Thorne~\cite{thorne74} that there is an upper limit  on the spin of a BH ($a/M\sim0.998$) through astrophysical processes like accretion. Similar results have been found in the context of more general conditions~\cite{klp10,nb18}. Motivated by results like Thorne's limit, we extend here our previous numerical work~\cite{bks21} to verify the existence of the Aretakis charge and Ori hair for near-extremal BHs i.e., BHs that deviate from extremality only very slightly ($1-q/M\sim10^{-9}$ -- $10^{-7}$ or $1-a/M\sim10^{-8}$ -- $10^{-4}$). Such small deviations from extremality will prove useful in our numerical analysis to demonstrate an approximately conserved Aretakis charge on the outer horizon, which has a potentially measurable signature outside the BH for relatively long times after an initial quasi-normal burst.\\

In this paper, we numerically demonstrate that the notion of the Aretakis scalar and gravitational hair extends to the case of near-extremal black holes as well, although as a {\em transient} effect. This paper is organized as follows: In Section II we establish the notation we use in this work and also offer details on coordinates and the numerical calculations. In Section III we present our results in the context of scalar field in NERN black hole spacetimes. In Section IV we focus on the most astrophysically relevant case of gravitational field in NEK spacetimes. We conclude with Section V and also mention planned future work. 

\section{Ori's Late-time expansion, Coordinates and Numerical Details}
We borrow our notation from previous work~\cite{bks21} where the late time expansion of a field $\Psi$ in a black hole space-time was written as:
\bea\label{ori_expansion}
    \Psi_{s,\ell,m}(t,r,\theta) =&& e^{I}_{s,\ell,m} r(r-M)^{-p^I_{s,\ell,m} } t^{-n^I_{s,\ell,m} } \Theta^I_{s,\ell,m}(\theta)\nn\\
    &&+\mathcal{O}(t^{-n^I_{s,\ell,m}-k^I_{s,\ell,m} })
\eea
in Boyer-Lindquist coordinates, where $s$ is the field's spin-weight, ($\ell, m$) characterize the spherical harmonic modes and the index $I$ on the right-side corresponds to the BH type i.e., $I = \{\text{NERN, NEK}\}$. We seek to verify the validity of this Ori-formula late time expansion of the field in the context of near-extremal BHs. Furthermore, we compare this Ori-coefficient with the Aretakis ``charge'' derived from the field on $\ml{H}^+$ and as done previously~\cite{bks21}, we will identify it as a (transient) ``hair'' since it is (nearly) constant and in-principle observable externally.\\

To do this, we solve the Teukolsky equation for perturbations in NERN and NEK BH backgrounds, focusing on axisymmetric modes ($m=0$). We modify the equation to work in compactified hyperboloidal coordinates $(\tau, \rho, \theta, \phi)$ that allow for time evolution on hypersurfaces which bring $\ml{I}^+$ to a finite radial coordinate $\rho(\ml{I}^+)=S<\infty$. The relationship between these new coordinates $(\tau,\rho)$ and the spherical Boyer-Lindquist coordinates $(t,r)$ is~\cite{bkz16}
\bea
\Omega &=& 1-\frac{\rho}{S}\nn\\
r &=& \frac{\rho}{\Omega(\rho)}\\
v\equiv t+r_*-r &=& \tau+\frac{\rho}{\Omega(\rho)}-\rho-4M\log\Omega(\rho)\nn
\eea
where $S$ denotes the location of $\ml{I}^+$ in hyperboloidal coordinates, $r_*$ is the usual tortoise coordinate and $v$ is the modified advanced time. Note that the angular variables are the same in both coordinate systems.\\

Our numerical implementation scheme entails re-writing the second order PDE in terms of two coupled first-order differential equations. We solve this system using a high-order WENO finite-difference scheme with explicit Shu-Osher time-stepping. Details may be found in our previous work~\cite{camc21}. We choose $S=19.0$ and the location of $\ml{H}^+$ such that for a similar mass EBH, $\rho_{EBH}(\ml{H}^+)=0.95$. The initial data is a truncated Gaussian centered at $\rho=1.0$ with a width of $0.22$ and non-zero for $\rho\in[0.95,8]$. This ensures compactly supported initial data but with non-zero support on the $\ml{H}^+$ surface. 

\section{Scalar Perturbations on RN Background}
\subsection{Aretakis charge and Ori expansion for scalar perturbations}
    We begin with solving the scalar equation in hyperboloidal coordinates on an RN background. This equation is separable into a radial-temporal $(1+1)$-D evolution equation and a stationary angular equation which is solved by the standard associated Legendre polynomials. The only angular dependence in the radial-temporal equation comes from the angular mode $\ell$ of the spherical harmonics. Below, we consider  two cases, $\ell=0,1$ on NERN backgrounds.\\

\begin{figure}[!ht]
    \centering
    \includegraphics[width=\columnwidth]{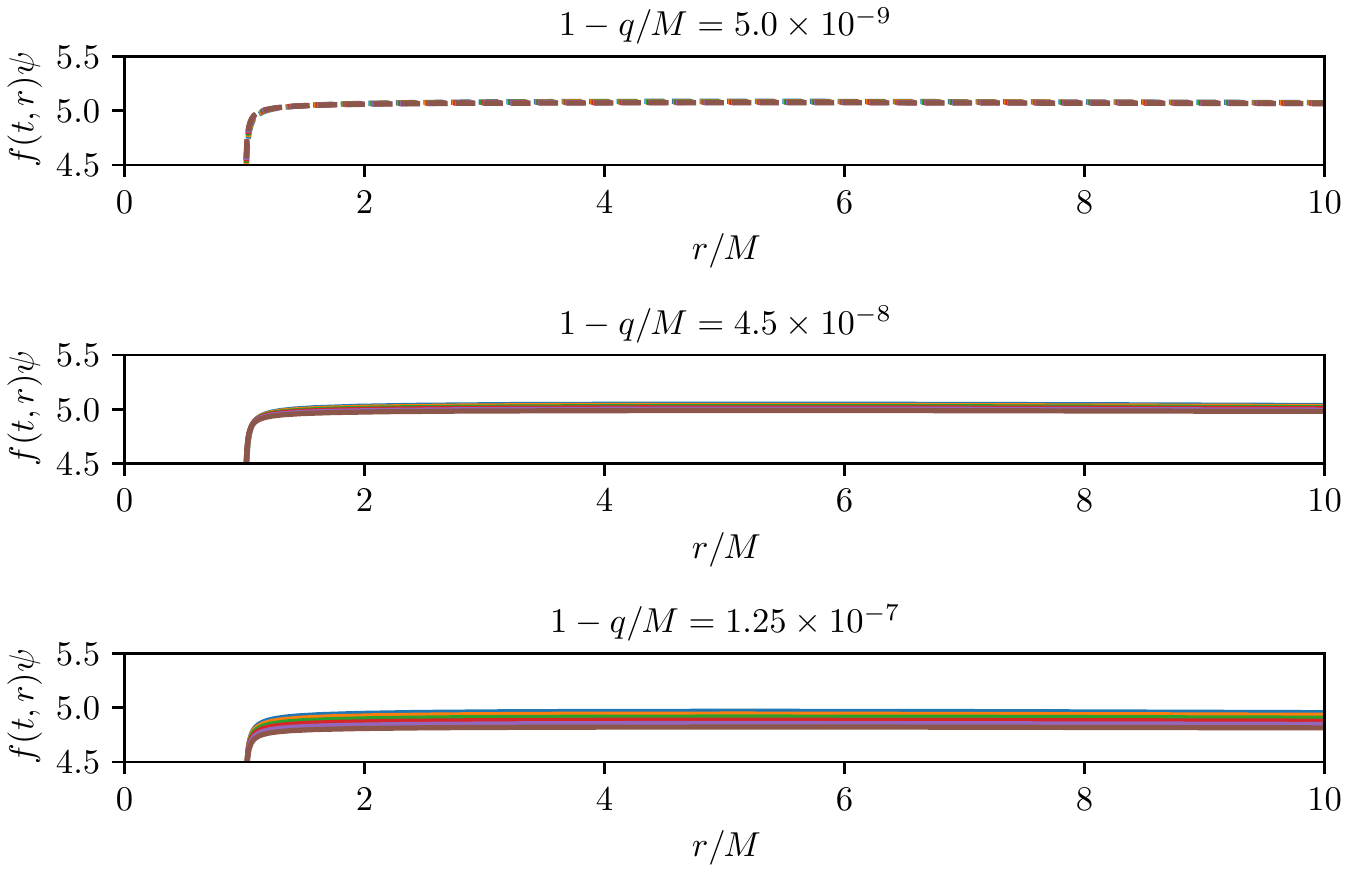}
    \caption{The values of $e^{RN}_{0,0,0}[\psi](t)$ as functions of $r/M$ for NERN.
These panels show values of the data set for which the Gaussian’s center is at $\rho/M = 1.0$. The values are plotted for $t/M =$ 1100, 1200, 1300, 1400, 1500, and 1600. The function $f(t, r) = (t/M)^2(1 -M/r)$.}
    \label{fig:Ori_RNL0}
\end{figure}

In Fig.~\ref{fig:Ori_RNL0}, we plot the function $\psi_{0,0,0}(t,r)f_{0,0,0}^{NERN}(t,r) = (t/M)^2(1 -M/r)\psi_{0,0,0}(t,r)$ as a function of $r/M$ for three different NERN BHs where $\psi$ is the scalar field. The initial condition is spherically symmetric ($\ell=0$) and so the product $\psi_{0,0,0}(t,r)f_{0,0,0}^{NERN}(t,r)$ corresponds to a pointwise estimation of $e_{0,0,0}^{NERN}$ with $p^{NERN}_{0,0,0}=1$ and $n^{NERN}_{0,0,0}=2$. We plot this product at five different times (as indicated in the caption) and note that the estimated Ori-coefficient appears to be a constant outside the BH. Since the angular component for a free massless scalar field equation is separable in an RN background and the initial data was rotationally invariant, we can immediately write down $\Theta^{NERN}_{0,0,0}=1.0$.

\begin{figure}[!ht]
    \centering
    \includegraphics[width=\columnwidth]{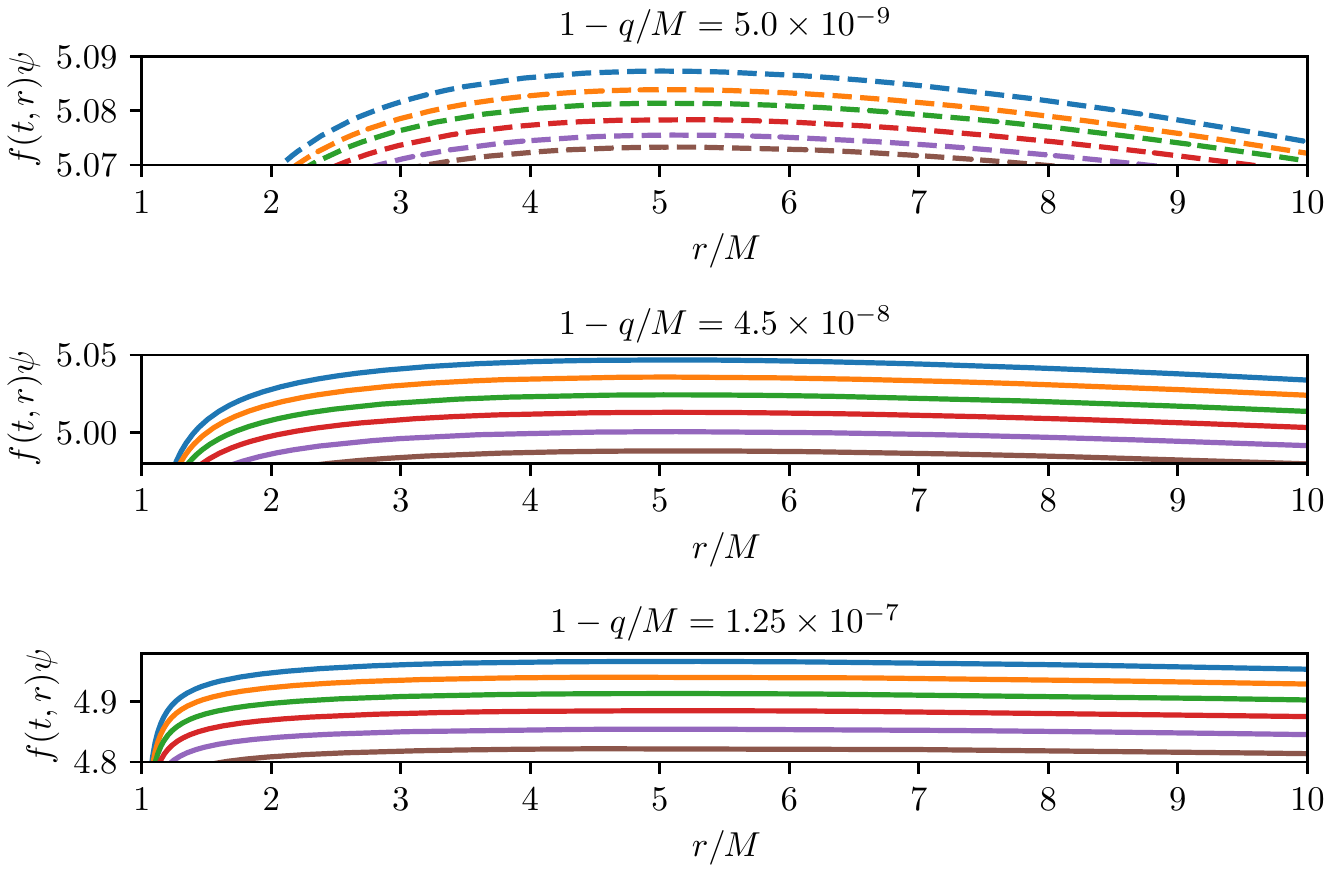}
    \caption{A zoomed in version of the estimated values of $e_{0,0,0}^{NERN}$ from Fig. \ref{fig:Ori_RNL0}.}
    \label{fig:Ori_RNL0_zoomed}
\end{figure}

At late times, the value of $\psi_{0,0,0}(t,r)f_{0,0,0}^{NERN}(t,r)$ asymptotes towards an approximate constant outside the BH. Moreover, as we drift away from extremality (top to bottom in Fig. \ref{fig:Ori_RNL0}), the value of $e_{0,0,0}^{NERN}$ and the field itself at a fixed $r$ and $t$ reduces (this is better illustrated in Fig. \ref{fig:Ori_RNL0_zoomed}) implying that the field decays slightly faster the more non-extremal the BH (all else being equal).

\begin{figure}[!ht]
    \centering
    \includegraphics[width=\columnwidth]{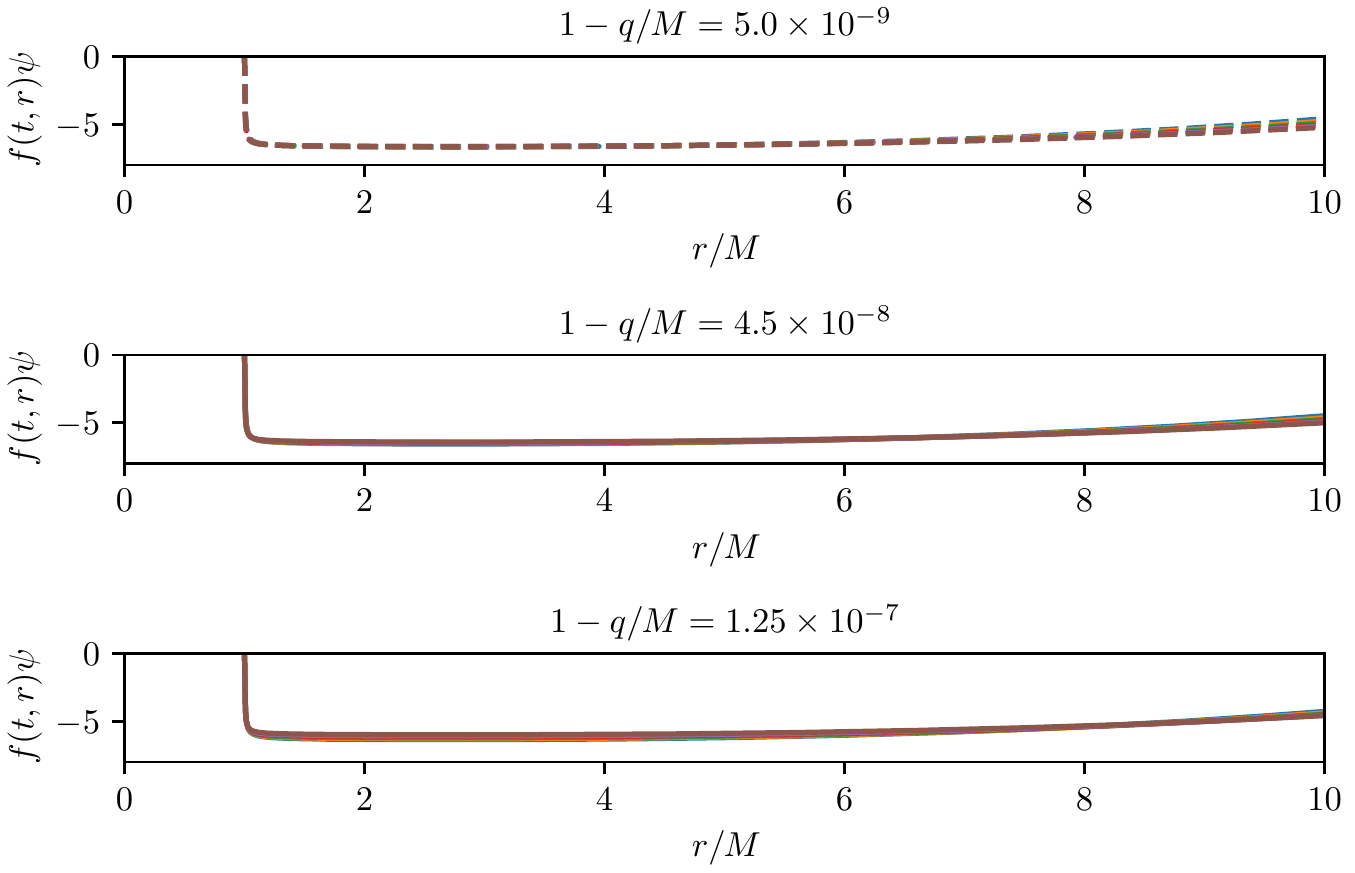}
    \caption{The values of $e^{NERN}_{0,1,0}[\psi](t)$ as functions of $r/M$ for NERN.
These panels show values of the data set for which the Gaussian’s center is at $\rho/M = 1.0$. The values are plotted for $t/M =$ 1100, 1200, 1300, 1400, 1500, and 1600. The function $f(t, r) = (t/M)^4(r/M)(1 -M/r)^2$.}
    \label{fig:Ori_RNL1}
\end{figure}

\begin{figure}[!ht]
    \centering
    \includegraphics[width=\columnwidth]{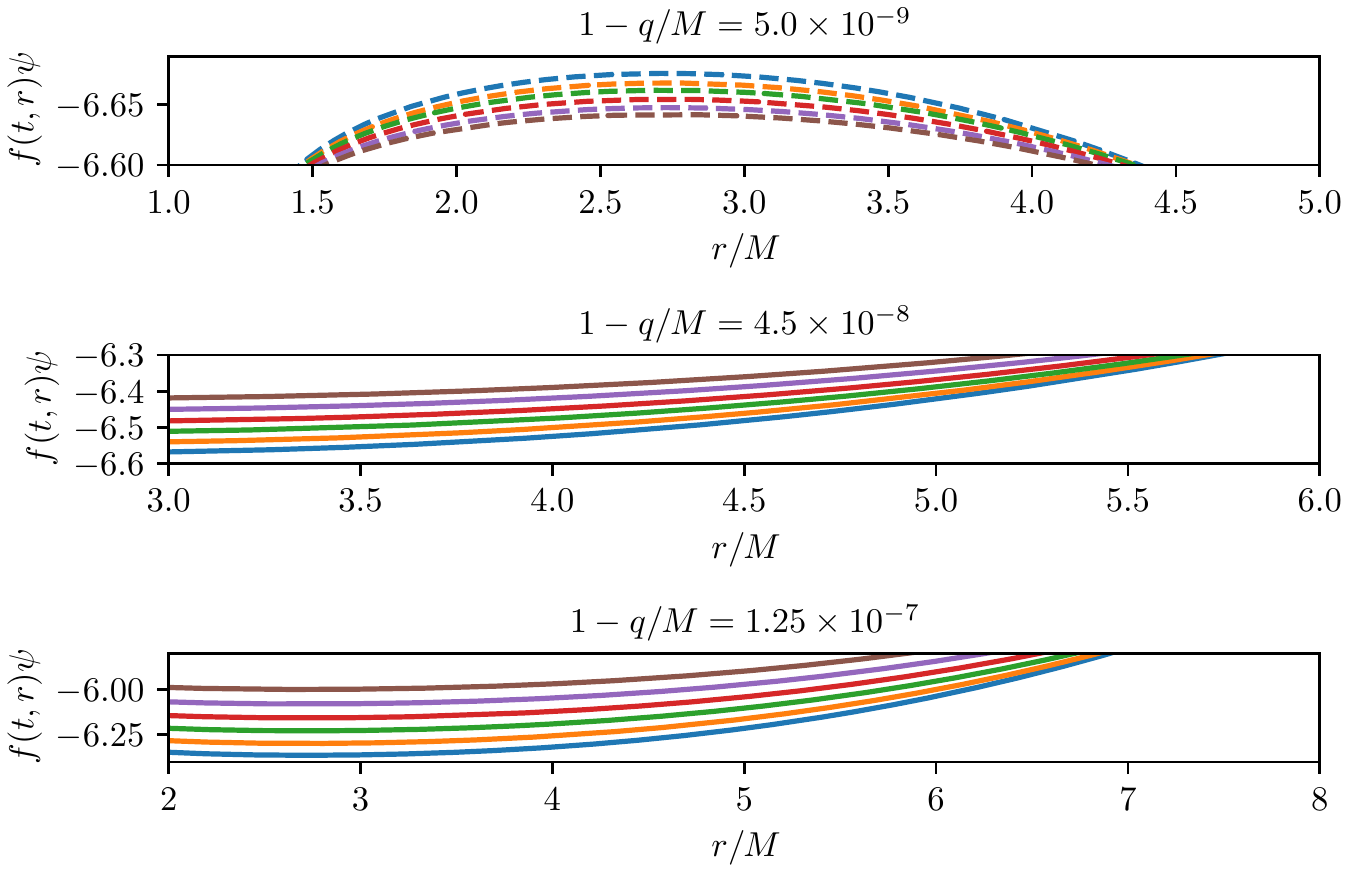}
    \caption{A zoomed in version of the estimated values of $e_{0,1,0}^{NERN}[\psi](t)$ from Fig. \ref{fig:Ori_RNL1}.}
    \label{fig:Ori_RNL1_zoomed}
\end{figure}

We continue our study to $\ell=1$ in NERN backgrounds in Fig.~\ref{fig:Ori_RNL1} which plots the radial evolution of $\psi_{0,1,0}(t,r)f_{0,1,0}^{NERN}(t,r)$, where $f_{0,1,0}^{NERN}(t,r)=(t/M)^4(r/M)(1 -M/r)^2$. Thus, this corresponds to a point-wise estimation of $e_{0,1,0}^{NERN}$ with $p^{NERN}_{0,1,0}=2$ and $n^{NERN}_{0,1,0}=4$. It is clear that for small values of $r$, the Ori expansion works well for NERN BHs even in the case of non-spherically symmetric perturbations, like $\ell=1$. However, as evident from Fig. \ref{fig:Ori_RNL1_zoomed}  we have a disagreement from the Ori expansion further from the BH. This is likely due to two reasons - (i) the field has not fully settled to its final asymptotic form yet or (ii) the nature of hyperboloidal slicing which tends towards the null direction at larger $\rho$ values.

\subsection{Transient nature of scalar hair for NERN}
\begin{figure}[!ht]
    \centering
    \includegraphics[width=\columnwidth,scale=0.8]{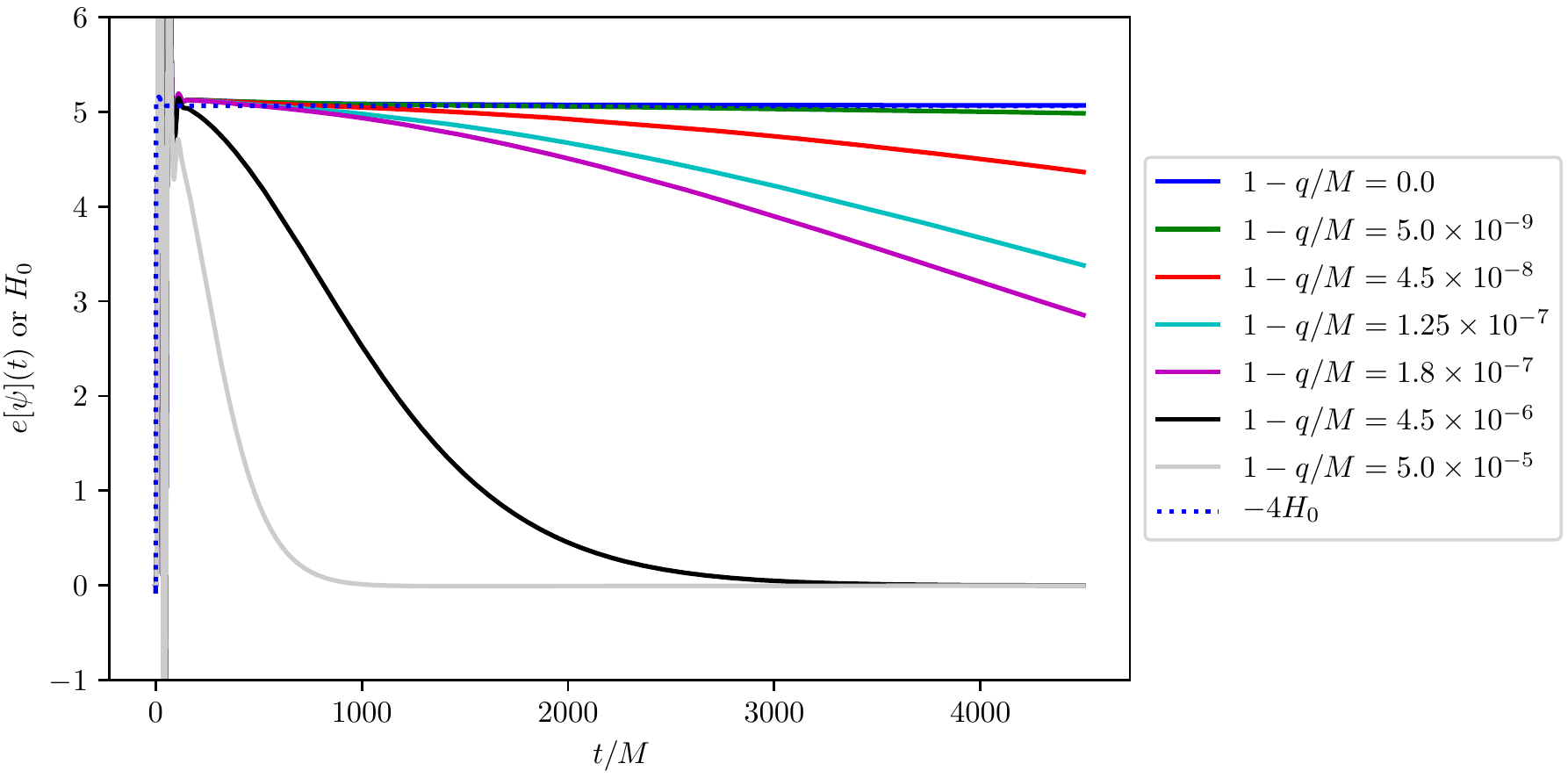}
    \caption{The values of $e^{NERN}_{0,0,0}[\psi](t)$ as functions of $t/M$. For every near-extremal $q$ value, a transient hair is observed. {\color{blue}}}
    \label{fig:OriL0_time}
\end{figure}

\begin{figure}[!ht]
    \centering
    \includegraphics[width=\columnwidth,scale=0.8]{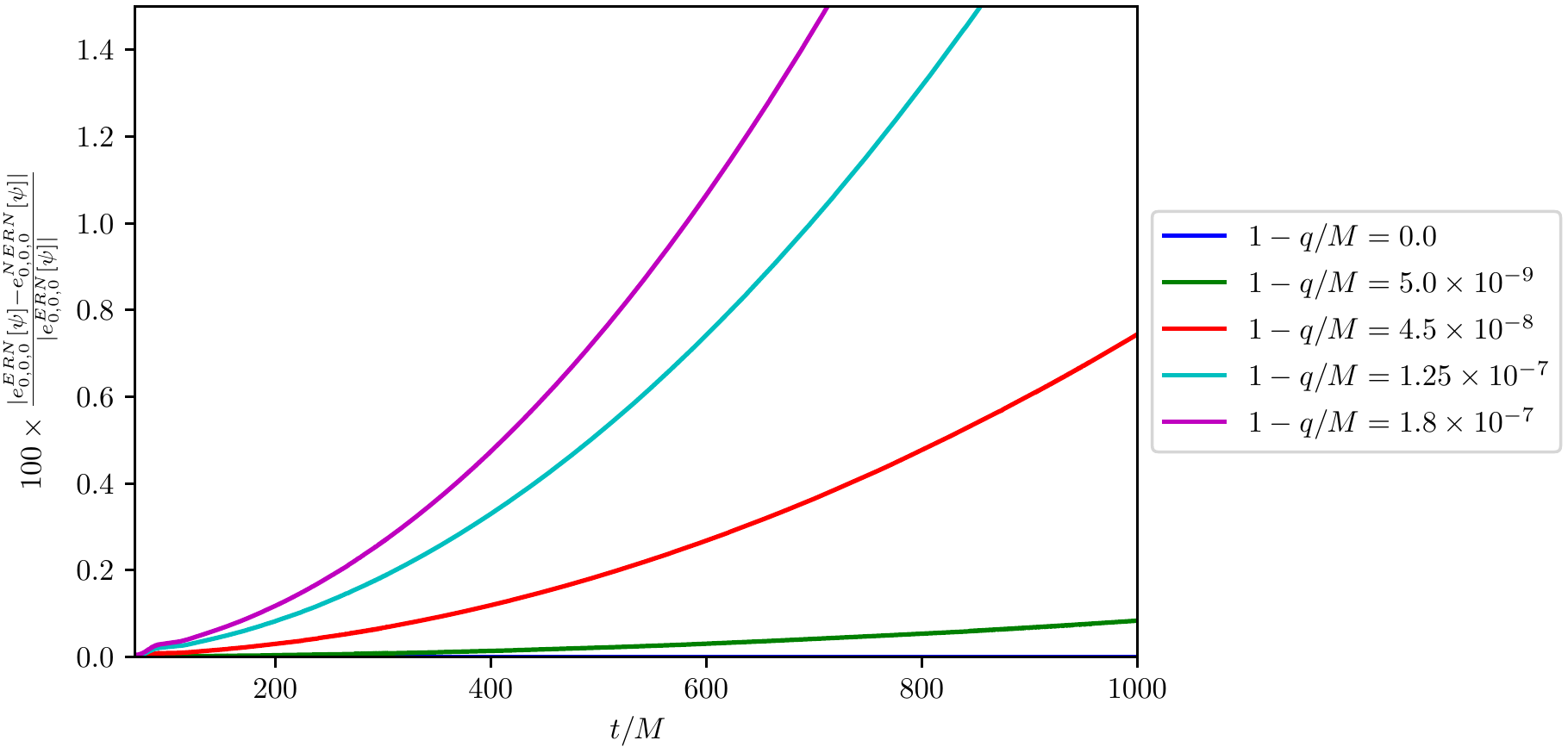}
    \caption{Percent difference between $e^{ERN}_{0,0,0}[\psi](q_{extremal}=M)$ and $e^{NERN}_{0,0,0}[\psi](q<M)$ for RN BH's and spherically symmetric ($\ell=0$) scalar perturbations.}
    \label{fig:PercentL0}
\end{figure}

To check the validity of the Ori expansion for the near-extremal cases and to estimate the temporal scale of validity, we plot in Fig.~\ref{fig:OriL0_time} the evolution of the estimated value of the Ori-coefficient $e^{NERN}_{0,0,0}$ as a function of time. As a reference, we also show the value of the Aretakis charge for the extremal case with the same initial perturbation. The plot shows a clear pattern - the further we depart from extremality (higher $1-q/M$), the smaller the time scale over which the Ori expansion or conserved Aretakis charge is prevalent. In addition, this conclusion is continuous in time i.e. {\em for arbitrarily small departures from extremality, the hair (manifested in Ori's coefficient, $e$) lasts for an arbitrarily long time}. In our most extremal case of $1-q/M=5.0 \times 10^{-9}$, we see divergence from the extremal case starting to appear around $t/M\sim3000-4000$. We make this concrete in Fig.~\ref{fig:PercentL0}, which plots the percent difference between the Ori's $e$ for a NERN BH compared to its value on an extremal RN background. In the case of $1-q/M=5.0 \times10^{-9}$ for example, the value of $e^{NERN}_{0,0,0}\left(1-q/M=5.0\times10^{-9}\right)$ is different from $e^{ERN}_{0,0,0}$ by only $\sim0.1\%$ at $t/M\sim1000$. Thus, even though the RN BH is non-extremal and the Aretakis charge (or its observable manifestation as Ori's $e$) is a transitory dynamical effect and if we wait long enough (large $t/M$) it would disappear (decay to zero), nonetheless it may last a long time depending on the severity of the BH's departure from extremality. Note that this observation was made by us previously in recently published work for the scalar field case using a different approach~\cite{bks19}.\\

\begin{figure}[!ht]
    \centering
    \includegraphics[width=\columnwidth,scale=0.8]{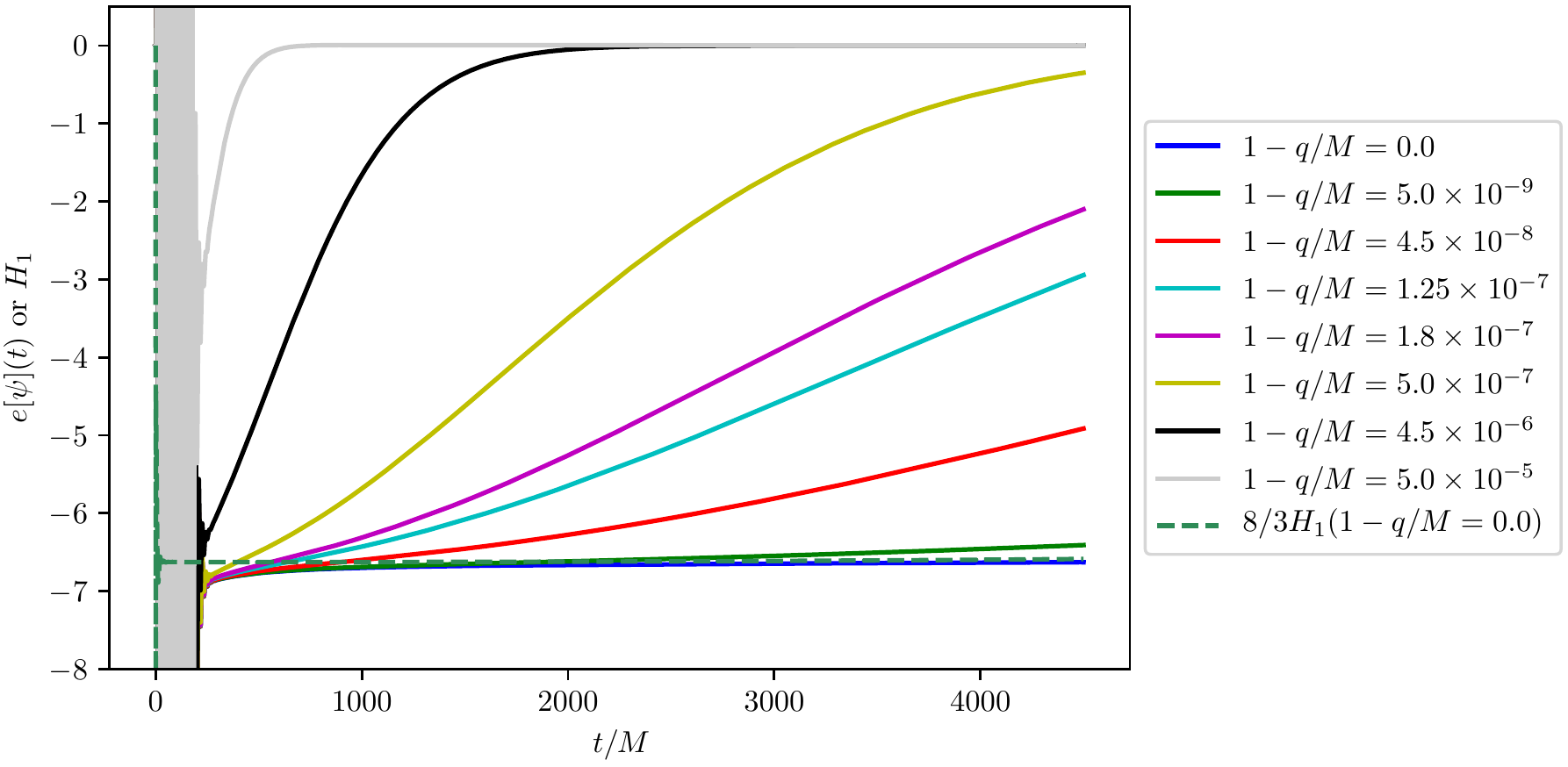}
    \caption{The values of $e^{NERN}_{0,1,0}[\psi](t)$ as functions of $t/M$ for NERN. For every near-extremal $q$ value, a transient hair is observed. {\color{blue}}}
    \label{fig:OriL1_time}
\end{figure}

\begin{figure}[!ht]
    \centering
    \includegraphics[width=\columnwidth,scale=0.8]{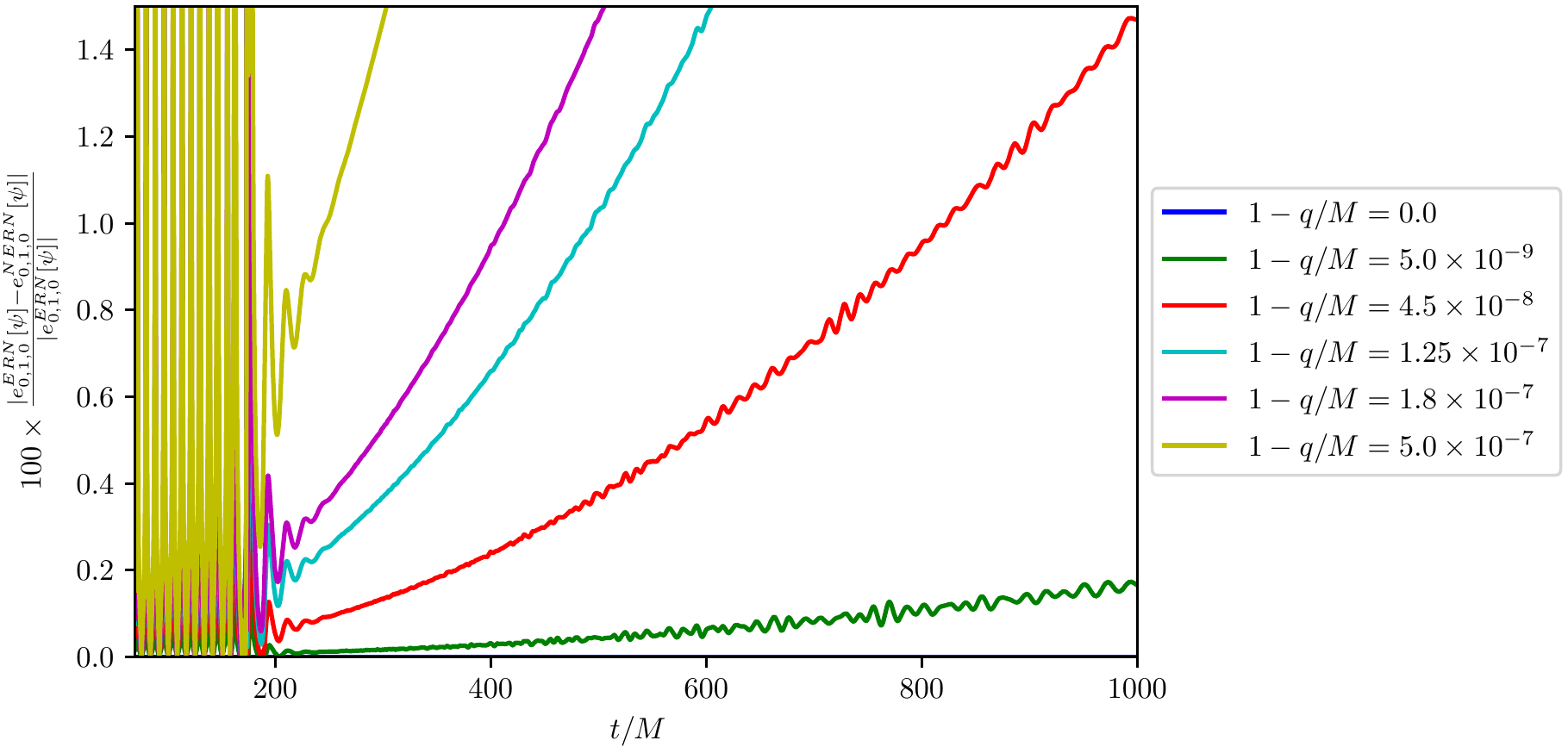}
    \caption{Percent difference between $e^{ERN}_{0,1,0}[\psi](q=M)$ and $e^{NERN}_{0,1,0}[\psi](q)$ for $\ell=1$ scalar perturbations.}
    \label{fig:PercentL1}
\end{figure}

We see in Figs.~\ref{fig:OriL1_time}, \ref{fig:PercentL1} that the above  conclusions about the transient nature of the hair for NERN BH's are robust for more general perturbations, like those with $\ell=1$. For example, in the case of $1-q/M=5\times10^{-9}$, even though the percent difference between $e^{NERN}_{0,1,0}$ and $e^{ERN}_{0,1,0}$ at $t/M\sim1000$ is a little higher than the $\ell=0$ case, it is still very small ($\sim0.2\%$). Figs.~\ref{fig:OriL0_time} and \ref{fig:OriL1_time} also include the dynamical evolution of the hair $e$ for the case of an extremal BH (with identical initial data as the near-extremal cases) for reference and we observe that this hair is a constant in the ERN case.\\

\begin{figure}[!ht]
    \centering
    \includegraphics[width=\columnwidth,scale=0.8]{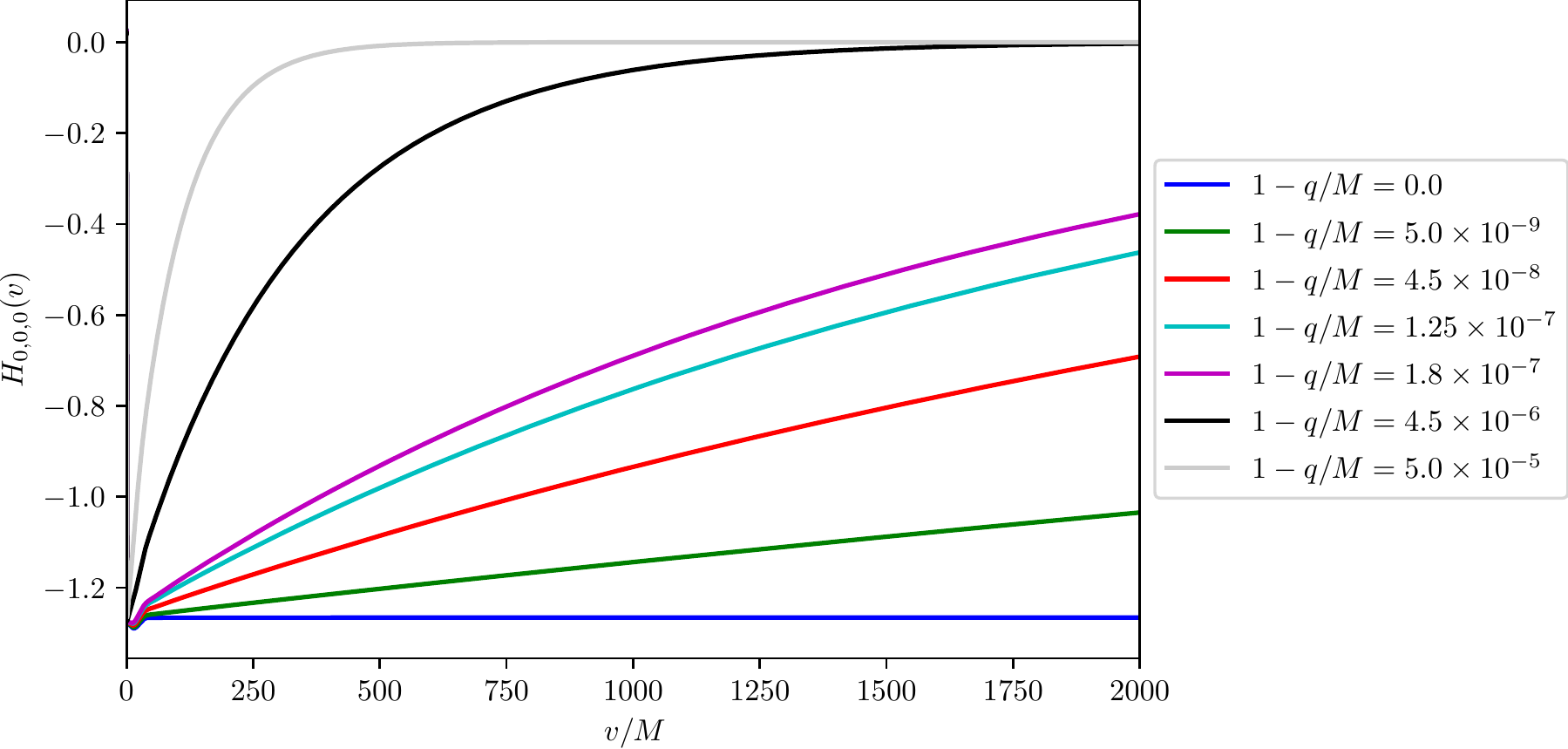}
    \caption{Aretakis charge, $H_{0,0,0}(v)$ calculated from the horizon integral for various NERN ($1-q/M$) as a function of the modified advanced time, $v$ for $\ell=0$ perturbation.}
    \label{fig:HorizonIntegralsL0}
\end{figure}

\begin{figure}[!ht]
    \centering
    \includegraphics[width=\columnwidth,scale=0.8]{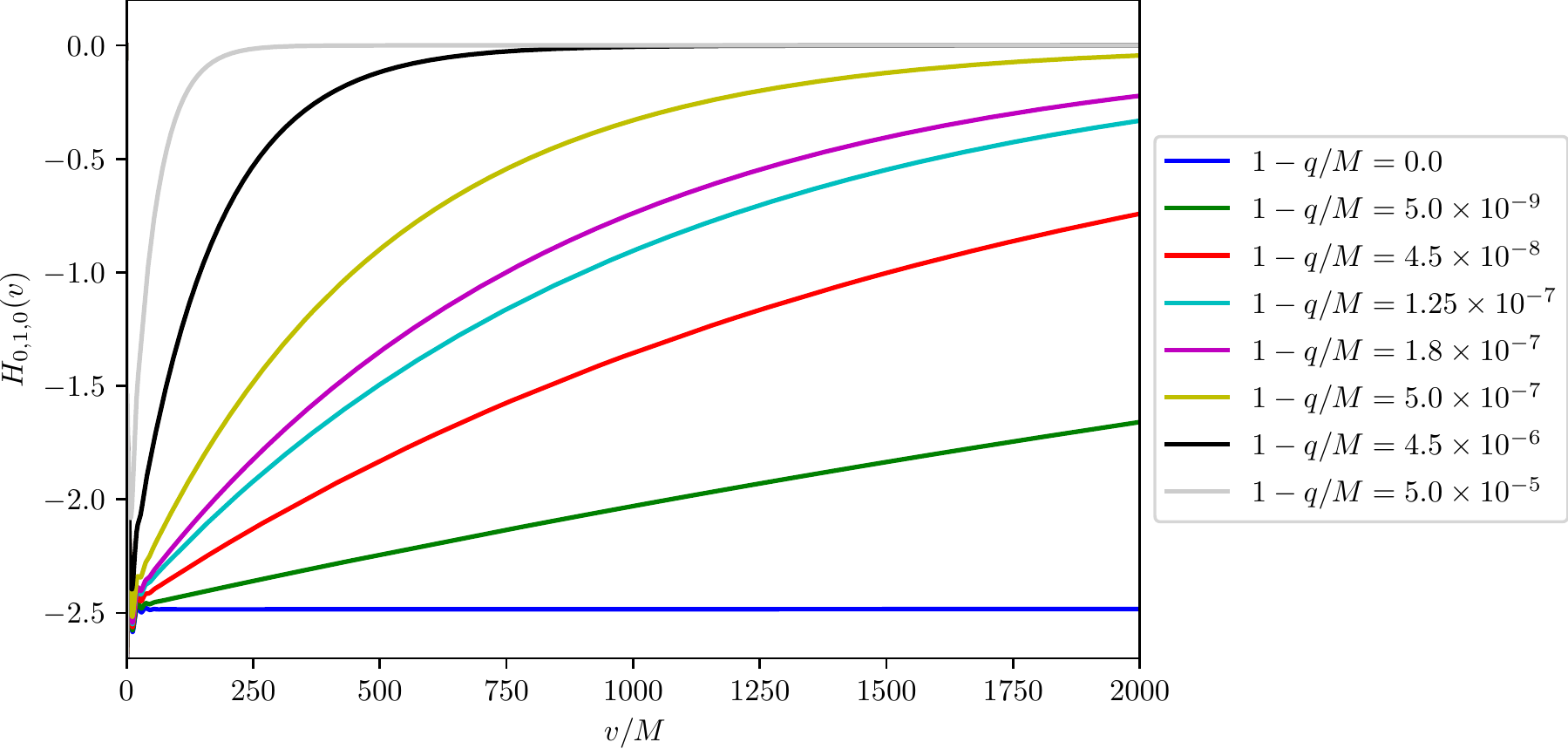}
    \caption{Aretakis charge, $H_{0,1,0}(v)$ calculated from the horizon integral for various NERN ($1-q/M$) as a function of the modified advanced time, $v$ for $\ell=1$ perturbation.}
    \label{fig:HorizonIntegralsL1}
\end{figure}

In Fig.~~\ref{fig:HorizonIntegralsL0}, \ref{fig:HorizonIntegralsL1} we plot the dynamical evolution of $H_{0,0,0}$, $H_{0,1,0}$ as functions of the modified advanced time $v$ on the horizon. These plots again indicate that the decay of the Aretakis charge is monotonic with $q/M$ -- the less extremal the BH (higher $1-q/M$), the faster the Aretakis charge decays to zero. This is in agreement with the monotonicity observed in the time evolution of the respective Ori constants in Figs.~\ref{fig:OriL0_time},~\ref{fig:OriL1_time}. Note that since neither the Aretakis charge nor the Ori-coefficient are precisely conserved for near-extremal BHs, we cannot compare them directly like we did in Ref.~\cite{bks21}. This is because of the different notions of time on the horizon and outside the BH.\\

We did not attempt additional simulations with higher values of $\ell$ due to the fact that those involve power-law ``tails'' that decay much faster resulting in the computational results accumulating too much floating-point error from round-off. This is a well-known challenge associated with numerical computations of radiative tails which is usually overcome using high-precision floating-point operations such as quadruple or octal precision~\cite{sk13}.

\section{Gravitational Perturbations on Near Extremal Kerr}
In the previous section, we considered scalar perturbations on NERN backgrounds and found that that there is an approximate conserved Aretakis charge observable (in principle) from outside the BH (in terms of Ori's constant, $e$) for late times. In this section, we extend this to gravitational perturbations on near-extremal Kerr BHs. Owing to their generic characteristics (like $a>0$ and $a/M<1$), such BHs are the most likely contenders for astrophysically relevant BHs.

\subsection{Aretakis charge and Ori expansion for gravitational perturbations}
To extract gravitational hair data outside the BH, we solve the Teukolsky equation with $s=-2$ (outgoing radiation) for a range of $a/M$ values, where $a$ is the Kerr spin parameter of the BH. In this work, we concentrate on the axisymmetric mode $m=0$, and so the dynamical system to solve is $(2+1)$-D in $(\tau,\rho,\theta)$. We denote the spin $2$ gravitational field by $\Phi$.\\

\begin{figure}[!ht]
    \centering
    \includegraphics[width=\columnwidth]{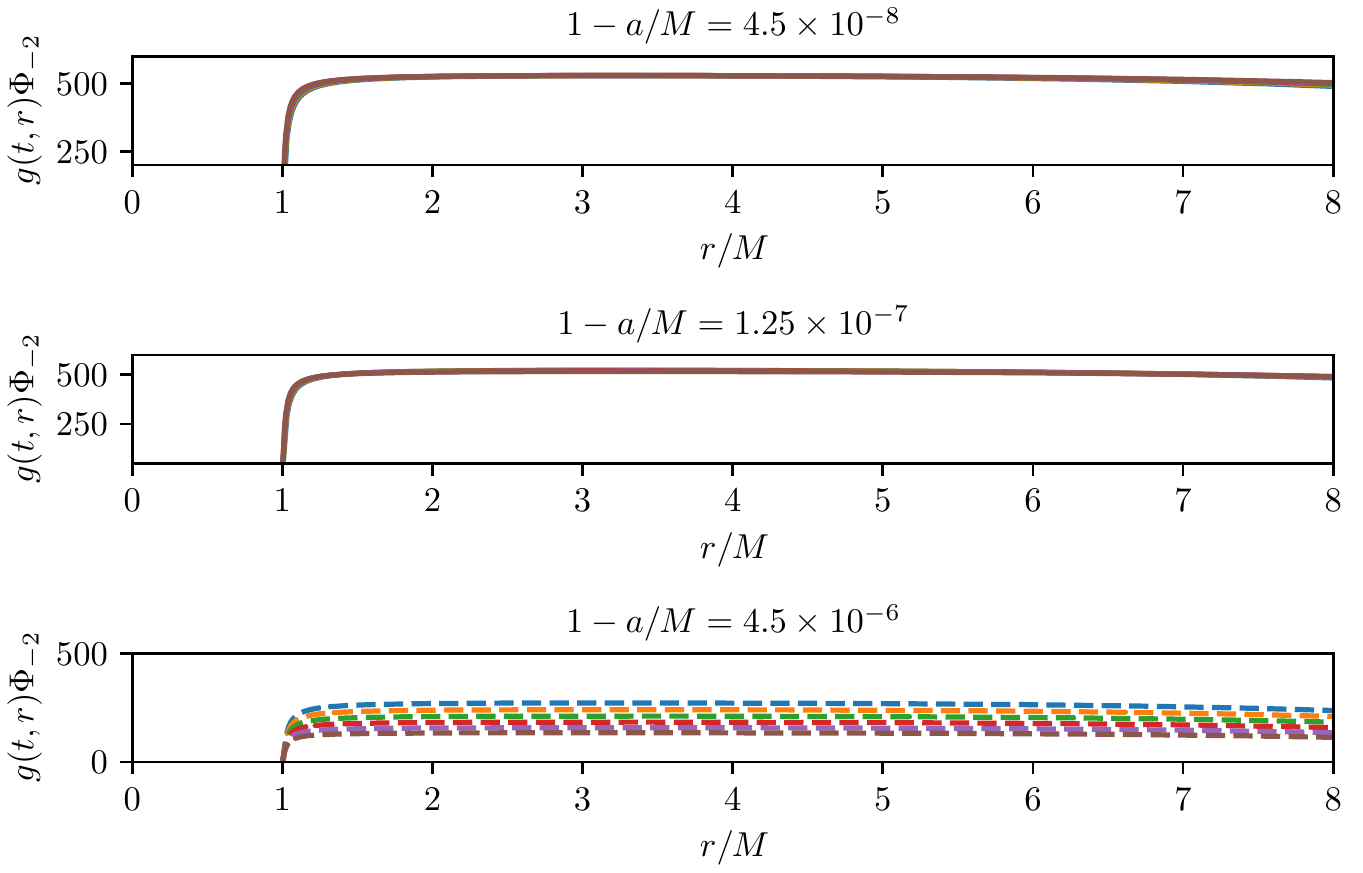}
    \caption{The values of $e^{NEK}_{-2,2,0}[\Phi](t)$ as functions of $r/M$ for NEK.
These panels show values of the data set for which the Gaussian’s center is at $\rho/M = 1.0$. The values are plotted for $t/M =$ 1100, 1200, 1300, 1400, 1500, and 1600. The function $g(t, r) = M(t/M)^6(r/M)^4(1 -M/r)^5$.}
    \label{fig:OriPhi_r}
\end{figure}

\begin{figure}[!ht]
    \centering
    \includegraphics[width=\columnwidth]{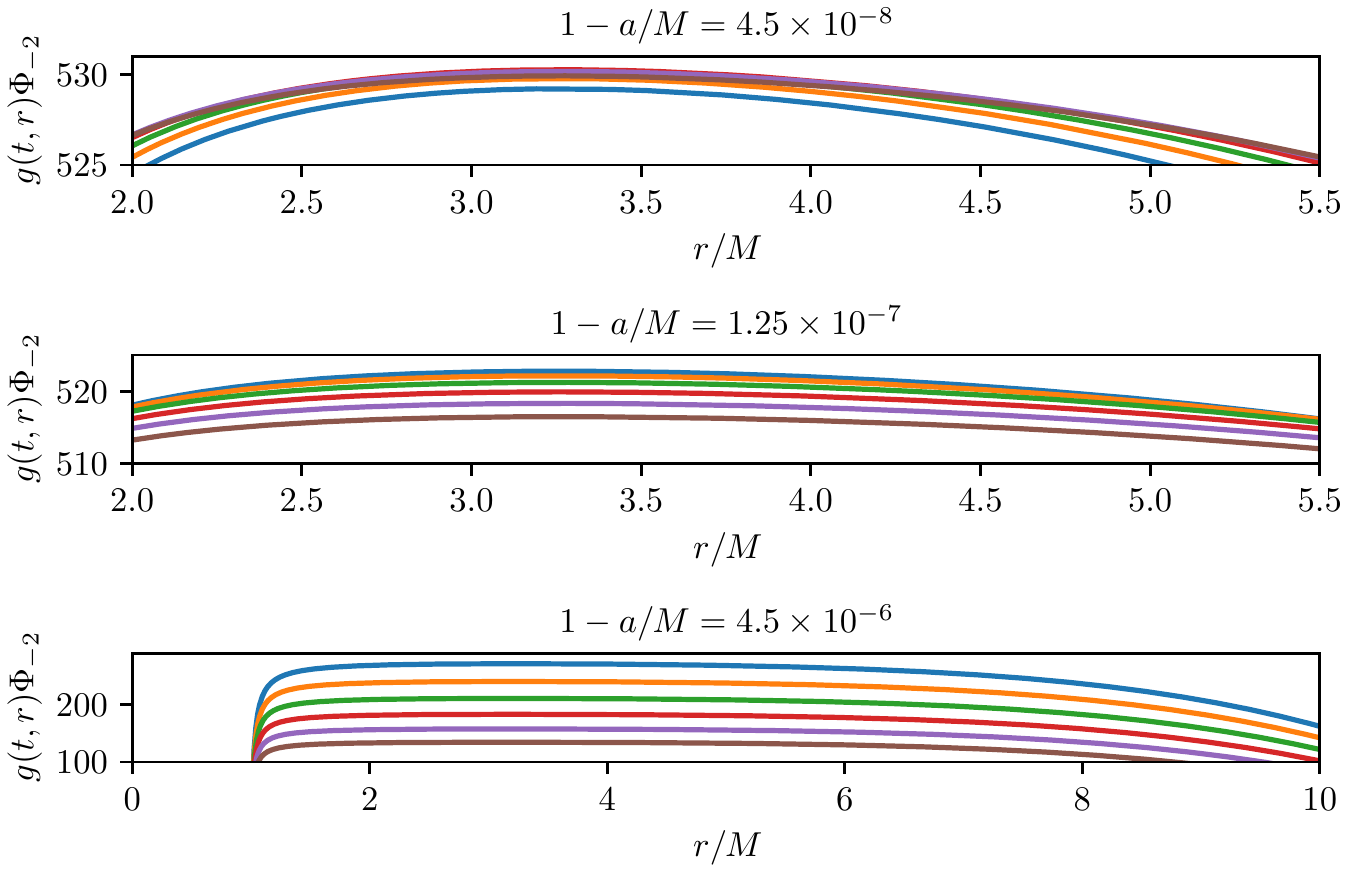}
    \caption{A zoomed in Fig. \ref{fig:OriPhi_r_zoomed} for $e^{NEK}_{-2,2,0}[\Phi](t)$.}
    \label{fig:OriPhi_r_zoomed}
\end{figure}

In Fig.~\ref{fig:OriPhi_r}, we plot $g^{NEK}_{-2,2,0}(t,r)\Phi_{-2}(t,r)$ as a function of $r/M$ where $g(t, r) = M(t/M)^6(r/M)^4(1 -M/r)^5$, as a proxy estimate of $e^{NEK}_{-2,2,0}$. It is clear that this radial dependence is fairly robust for various times $t/M\in[1100, 1600]$, at least for small $r$. We note again that the deviation at large $r$, as evidenced clearly in Fig. \ref{fig:OriPhi_r_zoomed} is an artifact of our limited evolution times or the hyperboloidal slicing tending towards a null direction. Finally, as we drift from extremality (higher $1-a/M$), the estimated $e^{NEK}_{-2,2,0}$ (at a fixed $r$ and $t$) decreases. Thus, we expect that the (transient) gravitational Aretakis charge for NEK would decay faster as the BH becomes less extremal. We show this explicitly in the next section.

\subsection{Transient nature of gravitational hair for NEK}
\begin{figure}[!ht]
    \centering
    \includegraphics[width=\columnwidth,scale=0.8]{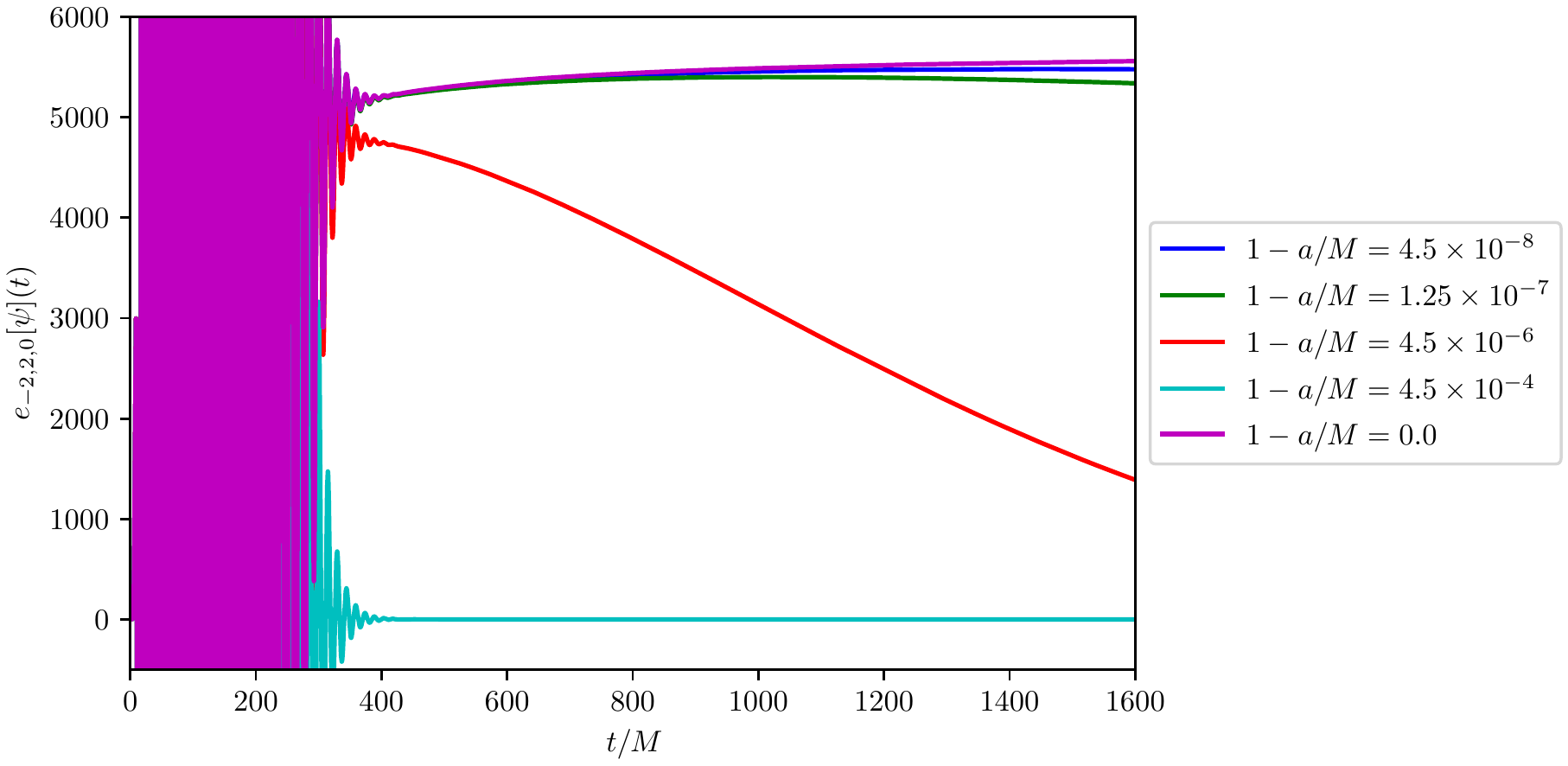}
    \caption{The values of $e^{NEK}_{-2,2,0}[\Phi](t)$ as functions of $t/M$ for NEK at $\rho/M=2.0$. }
    \label{fig:Ori_Phi_time}
\end{figure}

\begin{figure}[!ht]
    \centering
    \includegraphics[width=\columnwidth,scale=0.8]{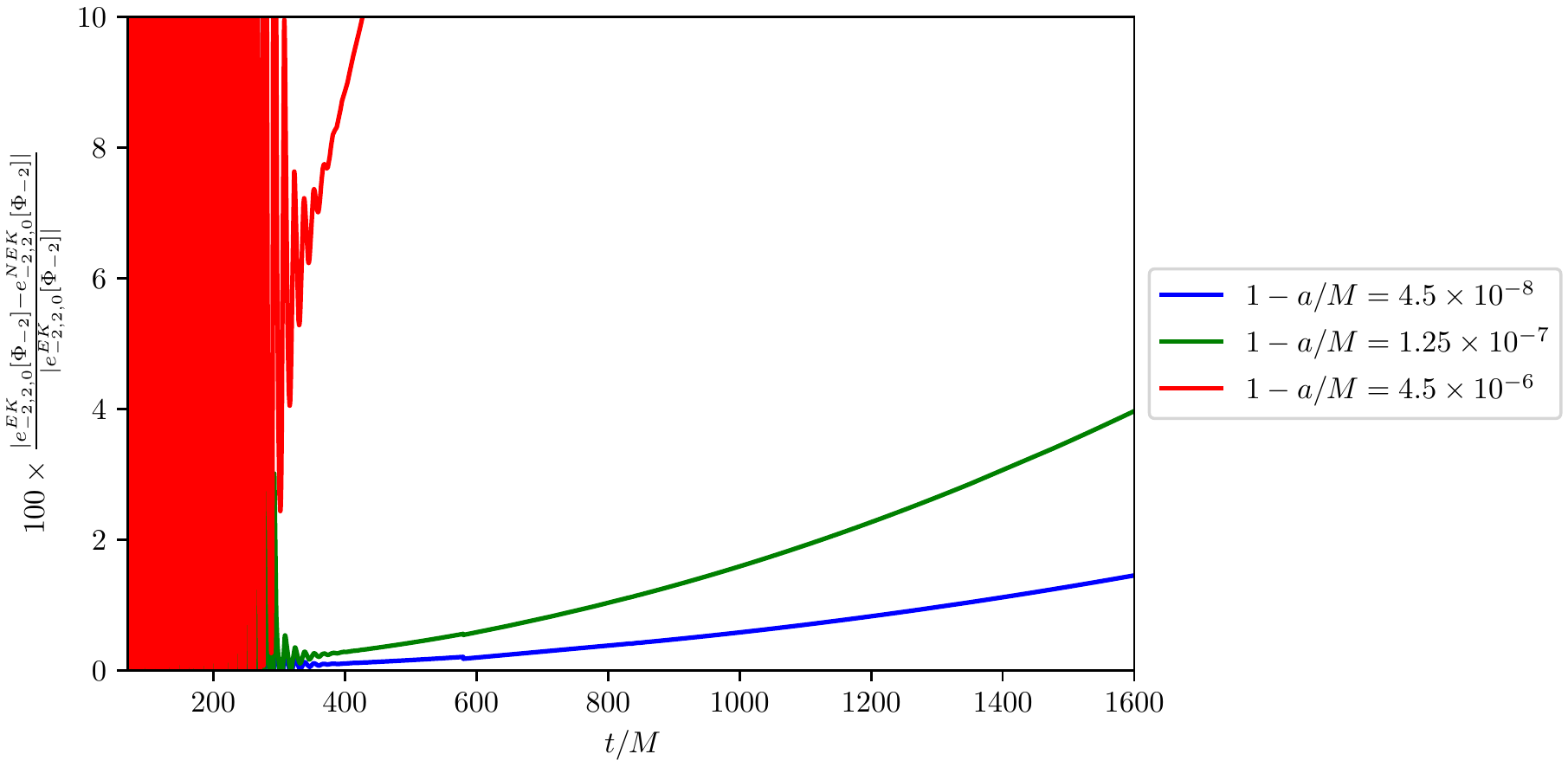}
    \caption{Percent difference between $e^{NEK}_{-2,2,0}[\Phi](t)$ and $e^{EK}_{-2,2,0}[\Phi](t)$ as functions of $t/M$ for NEK at $\rho/M=2.0$.}
    \label{fig:PercentL2}
\end{figure}

\begin{figure}[!ht]
    \centering
    \includegraphics[width=\columnwidth,scale=0.8]{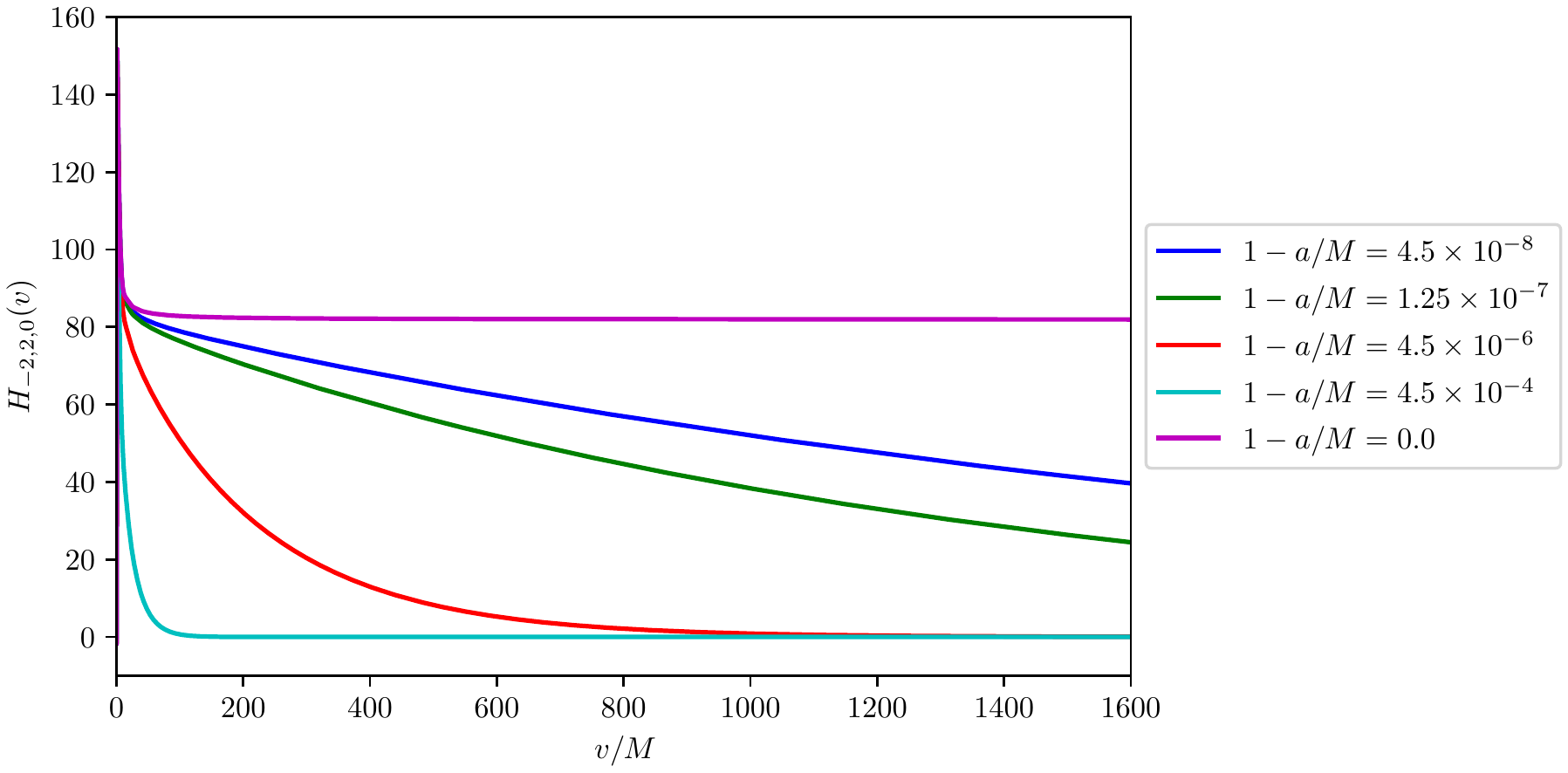}
    \caption{Aretakis charge, $H_{-2,2,0}(v)$ calculated from the horizon integral for varying $1-a/M$ NEK as a function of the modified advanced time, $v$. }
    \label{fig:H2}
\end{figure}

Figs.~\ref{fig:Ori_Phi_time} and \ref{fig:PercentL2} demonstrate the transient nature of the Ori-coefficient and the gravitational Aretakis charge. The Ori-coefficient remains constant for a relatively long time for NEK BHs and the closer we are to extremality (smaller $1-a/M$), the longer the ``conserved charge'' lingers. However, since the BH is non-extremal, the charge does start to decay at late times. We can see in Fig.~\ref{fig:Ori_Phi_time} that the estimated $e^{NEK}_{-2,2,0}$ at $r/M=2.0$ diverges from the extremal case (where it is a constant) at earlier times as the BH becomes less extremal (large $1-a/M$). In Fig.~\ref{fig:H2} we also show the evolution of the gravitational Aretakis charge $H_{-2,2,0}(v)$ as a function of $v$. We again note that as the BH becomes more extremal (smaller $1-a/M$), the charge $H_{-2,2,0}(v)$ decays slowly becoming exactly constant when the BH is extremal. In other words, {\em the closer the BH is to extremality, the longer the gravitational hair persists in value closer to its value in the extremal case, allowing for it to be potentially observable}.  

\section{Conclusions and Future Work}
In this paper, we numerically demonstrated that expressions for the Aretakis charge or the Ori coefficient $e$ in the context of extremal BHs can also be used for near-extremal cases. For the near-extremal cases, such quantities are not strictly conserved; however, they can be made to stay nearly constant for arbitrarily long duration depending on the charge or spin of the BH. We included results for the case of scalar perturbations ($\ell=0,1$) on NERN BH spacetimes. In our most extremal case ($1-q/M=5\times10^{-9}$) for example, we found that that the divergence of the Ori-coefficient from its extremal counterpart was very small $(\sim0.1-0.2\%$) at times as late as $\ml{O}(t/M)\sim1000$.\\

We also considered the astrophysically interesting case of gravitational perturbation on NEK BH spacetimes. Similar to the scalar case on NERN backgrounds, we found that the gravitational hair (as manifested in it is Ori-coefficient) is approximately preserved (constant) for a long time, even though it is not strictly conserved. In our most extremal case for Kerr ($1-a/M=4.5\times10^{-8}$), the Ori-coefficient diverged from its extremal counterpart by $\sim1\%$ for $\ml{O}(t/M)\sim1000$. Thus, such an astrophysical NEK could have detectable remnants of a gravitational hair. It is yet unclear if such high-spin NEK black holes are prevalent in nature.\\

We limited ourselves to axisymmetric field configurations on Kerr for this study. In future work, we plan to generalize the gravitational hair to the non-axisymmetric case and also perform a detailed detectability analysis in the context of gravitational wave observatories like LIGO-Virgo-KAGRA.\\

\noindent{\em Acknowledgements:} The authors thank Lior Burko for discussions. K. G-Q. is grateful for support from UMass Dartmouth and the Center for Scientific Computing \& Data Research and help from Tousif Islam on some of the figures. G.K. acknowledges research support from NSF Grants PHY-2106755 and DMS-1912716. All the computations were performed on the MIT Lincoln Labs {\em SuperCloud} GPU supercomputer supported by the Massachusetts Green High Performance Computing Center (MGHPCC).

\end{document}